\documentclass[aps,pra,reprint,amsmath,amsfonts,amssymb,superscriptaddress]{revtex4-1}
\usepackage{graphicx,calc}
\usepackage{color,bm}
\usepackage{braket}
\usepackage[colorlinks,urlcolor=blue,citecolor=blue,linkcolor=blue]{hyperref}
\usepackage{makecell}
\begin{document}

\title{Terahertz Receiver based on Room-Temperature Rydberg-Atoms}

\author{Yayi Lin}
\altaffiliation{Y. Lin and Z. She contributed equally to this work.}
\affiliation{Guangdong Provincial Key Laboratory of Quantum Engineering and Quantum Materials, School of Physics and Telecommunication Engineering, South China Normal University, Gua,.ngzhou 510006, China}
\affiliation{Guangdong-Hong Kong Joint Laboratory of Quantum Matter, Frontier Research Institute for Physics, South China Normal University, Guangzhou 510006, China}

\author{Zhenyue She}
\altaffiliation{Y. Lin and Z. She contributed equally to this work.}
\affiliation{Guangdong Provincial Key Laboratory of Quantum Engineering and Quantum Materials, School of Physics and Telecommunication Engineering, South China Normal University, Guangzhou 510006, China}
\affiliation{Guangdong-Hong Kong Joint Laboratory of Quantum Matter, Frontier Research Institute for Physics, South China Normal University, Guangzhou 510006, China}

\author{Zhiwen Chen}
\affiliation{Guangdong Provincial Key Laboratory of Quantum Engineering and Quantum Materials, School of Physics and Telecommunication Engineering, South China Normal University, Guangzhou 510006, China}
\affiliation{Guangdong-Hong Kong Joint Laboratory of Quantum Matter, Frontier Research Institute for Physics, South China Normal University, Guangzhou 510006, China}

\author{Xianzhe Li}
\affiliation{Guangdong Provincial Key Laboratory of Quantum Engineering and Quantum Materials, School of Physics and Telecommunication Engineering, South China Normal University, Guangzhou 510006, China}
\affiliation{Guangdong-Hong Kong Joint Laboratory of Quantum Matter, Frontier Research Institute for Physics, South China Normal University, Guangzhou 510006, China}

\author{Caixia Zhang}
\affiliation{Guangdong Provincial Key Laboratory of Quantum Engineering and Quantum Materials, School of Physics and Telecommunication Engineering, South China Normal University, Guangzhou 510006, China}
\affiliation{Guangdong-Hong Kong Joint Laboratory of Quantum Matter, Frontier Research Institute for Physics, South China Normal University, Guangzhou 510006, China}

\author{\\Kaiyu Liao}
\affiliation{Guangdong Provincial Key Laboratory of Quantum Engineering and Quantum Materials, School of Physics and Telecommunication Engineering, South China Normal University, Guangzhou 510006, China}
\affiliation{Guangdong-Hong Kong Joint Laboratory of Quantum Matter, Frontier Research Institute for Physics, South China Normal University, Guangzhou 510006, China}
\affiliation{GPETR Center for Quantum Precision Measurement, South China Normal University, Guangzhou 510006, China}

\author{Xinding Zhang}
\affiliation{Guangdong Provincial Key Laboratory of Quantum Engineering and Quantum Materials, School of Physics and Telecommunication Engineering, South China Normal University, Guangzhou 510006, China}
\affiliation{Guangdong-Hong Kong Joint Laboratory of Quantum Matter, Frontier Research Institute for Physics, South China Normal University, Guangzhou 510006, China}
\affiliation{GPETR Center for Quantum Precision Measurement, South China Normal University, Guangzhou 510006, China}
\affiliation{SCNU Qingyuan Institute of Science and Technology Innovation Co., Ltd., Qingyuan 511517, China}

\author{Wei Huang}
\email{WeiHuang@m.scnu.edu.cn}
\affiliation{Guangdong Provincial Key Laboratory of Quantum Engineering and Quantum Materials, School of Physics and Telecommunication Engineering, South China Normal University, Guangzhou 510006, China}
\affiliation{Guangdong-Hong Kong Joint Laboratory of Quantum Matter, Frontier Research Institute for Physics, South China Normal University, Guangzhou 510006, China}
\affiliation{GPETR Center for Quantum Precision Measurement, South China Normal University, Guangzhou 510006, China}

\author{Hui Yan}
\email{yanhui@scnu.edu.cn}
\affiliation{Guangdong Provincial Key Laboratory of Quantum Engineering and Quantum Materials, School of Physics and Telecommunication Engineering, South China Normal University, Guangzhou 510006, China}
\affiliation{Guangdong-Hong Kong Joint Laboratory of Quantum Matter, Frontier Research Institute for Physics, South China Normal University, Guangzhou 510006, China}
\affiliation{GPETR Center for Quantum Precision Measurement, South China Normal University, Guangzhou 510006, China}

\author{Shiliang Zhu}
\email{slzhu@scnu.edu.cn}
\affiliation{Guangdong Provincial Key Laboratory of Quantum Engineering and Quantum Materials, School of Physics and Telecommunication Engineering, South China Normal University, Guangzhou 510006, China}
\affiliation{Guangdong-Hong Kong Joint Laboratory of Quantum Matter, Frontier Research Institute for Physics, South China Normal University, Guangzhou 510006, China}

\begin{abstract}
Realization of practical terahertz wireless communications still faces many challenges. The receiver with high sensitivity is important for THz wireless communications. Here we demonstrate a terahertz receiver based on the cesium Rydberg atoms in a room-temperature vapor cell. The minimum detectable THz electric field is calibrated. With this receiver, the phase-sensitive conversion of amplitude-modulated or frequency-modulated terahertz waves into optical signals is performed. The results show that the atomic receiver has many advantages due to its quantum properties. Especially, the long distance THz wireless communications is achievable using this receiver. Furthermore, the atomic receiver can be used in the THz wireless-to-optical link.
\end{abstract}

\maketitle

\section{Introduction}
Since the data traffic requirement in wireless communications networks is increasing exponentially, the spectral band resources will inevitably be extended to terahertz (THz). And integration of THz links into fibre-optic infrastructures seamlessly is also crucial important \cite{SeedsJLT2015,YuAPLL2016,WangAPLL2020}. Significant progress has been made to construct the high performance THz wireless communications systems which mainly consist of transmitters and receivers\cite{KurnerJIMTW2014,LiuOE2005,ShamsOE2014,NagatsumaOE2013,LiOE2013,NagatsumaNPon2016,HarterNPon2020,HarterOptica2019,WangIEEE2014}. By combining the photonic transmitter and electronic receiver, the high data rate system transmitting data over 20 $m$  at 237.5 $GHz$ was realized \cite{KoenigNPon2013}. Besides, the graphene-based devices \cite{TongNL2015, LiuNC2015}, plasma-based transistors \cite{OtsujiTTST2015,TohmeEL2014} and topological photonics are also promising technologies for THz communications\cite{YangNP2020}. However, the practical applications of THz wireless communications systems still face great challenges. The severe attenuation makes long distance THz wireless communications difficult \cite{HanTTST2016,HanTWC2015}. The receiver with high sensitivity is important for THz wireless communications. Another side, the THz-over-fiber architecture using receiver which can directly convert received wireless signals into optical signals is also of great importance but technically difficult \cite{SalaminNP2018,UmmethalaNP2019}.

\begin{figure*}[ptb]
	\centering\includegraphics[width=18cm]{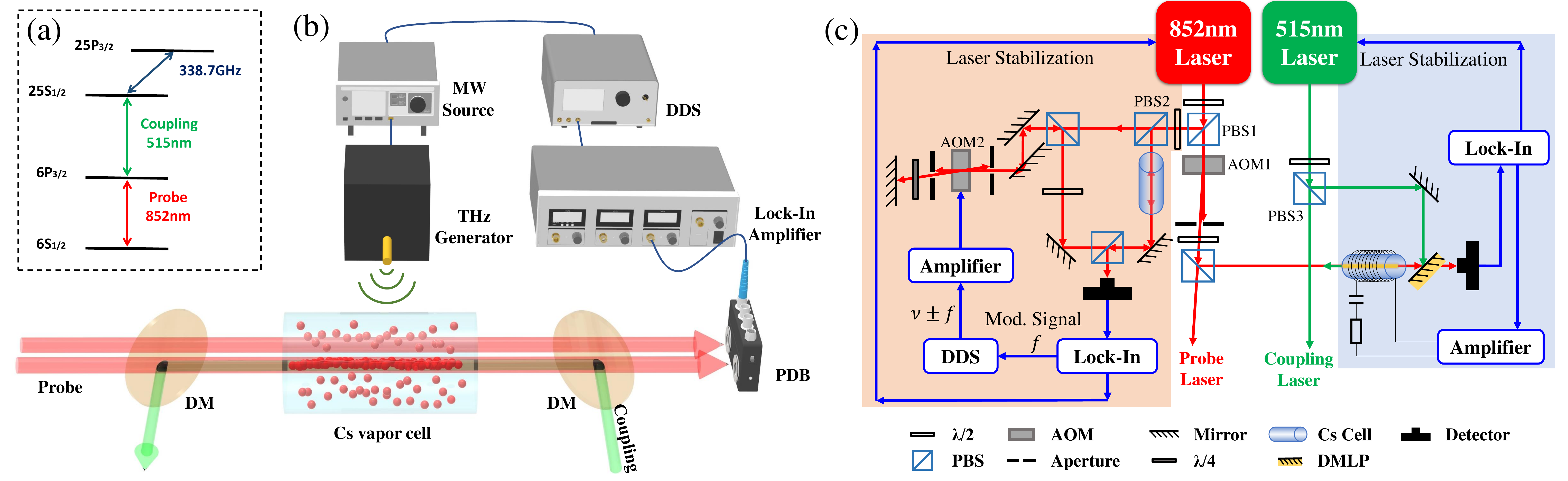}
	\caption{\label{fig1} Level diagram and experimental setup.
		(a) The four-level system used for the experiments. The probe and coupling light counter-propagate in the vapor cell of cesium atoms. Two dichroic mirrors (DM) are used to combine and split the two beams separately as experiment needed. The terahertz wave couples the Rydberg states $25S_{1/2}$ and $25P_{3/2}$. 
		(b) The experimental setup. The terahertz wave is generated by a 32-times frequency multiplier chain with a commercial microwave source. A two channel direct digital synthesis signal (DDS) generator is used to provide the phase modulated signal for THz wave and reference for lock-in amplifier. An amplified balanced photodetector (PDB) is used to detect the transmission of probe beam.  The communications information is demodulated by the lock-in amplifier.
        (c)The laser stabilization system by frequency modulation methods. The 852nm probe laser is frequency-locked with the modulation transfer spectroscopy (orange region) and the 515nm coupling laser is stabilized by Zeeman modulation method (blue region).}
\end{figure*}

Recently, the Rydberg atoms have been demonstrated as electromagnetic field sensors with ultrahigh sensitivity and capability of self-calibration \cite{SedlacekNP2012, HollowayJAP2017, JingNP2020,LiaoPRA2020,ChenAPS2021,HuangAPS2015}: the Rydberg-atom sensors have been proved to be excellent quantum receivers for the microwave communications \cite{MeyerAPL2018,SongOE2019,SimonsAPL2019}. In the THz frequency range, the Rydberg atoms still have large dipole moment making them ideal candidates for THz quantum sensors \cite{WadeNP2017,DownesPRX2020,WadeNC2018,MeyerJPB2020}. The ultimate performance of Rydberg-atom receiver is restricted by the quantum limits. This enables Rydberg-atom receiver having potential to realize long distance THz wireless communications. Meanwhile, the atomic receiver can be used in the THz wireless-to-optical link because it can directly convert the wireless signals into the optical ones \cite{DebAPL2018}.

In this paper, we demonstrate that the Rydberg atoms in a room-temperature cesium (Cs) vapor cell can be an excellent receiver for THz wireless signals. The minimum detectable THz electric field using electromagnetically induced transparency (EIT) and Autler-Townes splitting (ATS) technique is calibrated.  Then we implement the atomic receiver for THz wireless communications with a 32-times frequency multiplier source. The phase-sensitive conversion of amplitude-modulated or frequency-modulated THz wave into optical signals is used to perform a canonical digital communications with 8-state phase-shift-keying (PSK). Our work demonstrates that the THz receiver based on Rydberg-atoms has advantages due to its quantum properties. Especially, the long distance THz wireless communications is achievable using this receiver.

\section{Experimental setup}
The schematic of the standard four-level EIT-based Rydberg receiver is illustrated in Fig.~\ref{fig1}. The probe and coupling laser beams counter-propagate through a room-temperature Cs vapor cell. The probe laser is 100 $uW$ with $1/e^2$ radius 120 $\mu m$ and the coupling laser is 14.5 $mW$ with $1/e^2$ radius 70 $\mu m$. An amplified balanced photodetector (PDB) is served to measure the transmitted probe power subtracting the noisy background to observe EIT. A microwave source (R\&S SMF100A) and a 32-times frequency multiplier chain with 26 $dBi$ THz horn provides the 338.7 GHz wave to address the $25S_{1/2}\rightarrow$$25P_{3/2}$ Rydberg transition. This THz horn is 15 $cm$ away from the science cell.

To avoid interfering the probe transmission, both lasers are locked by frequency modulation methods using separated cells. The 852 nm probe beam is locked to the $6S_{1/2}(F=4)\rightarrow$$6P_{3/2}(F'=5)$ transition using the modulation transfer spectroscopy\cite{PreuschoffOE2018}. As shown in Fig.~\ref{fig1}(c), the frequency is modulated by AOM2 with $f=\pm1$ MHz. The error signal is demodulated through a Lock-In amplifier from the modulation transfer spectroscopy to stabilize the laser frequency. While the 515 nm coupling laser is locked to the $6P_{3/2}\rightarrow$$25S_{1/2}$ transition using the Zeeman modulation scheme\cite{JiaAO2020}. The coil cover the Cs vapour cell is used to apply the Zeeman modulation with frequency 27 $kHz$. The error signal can be obtained from the EIT signal. Both lasers can be stabilized with linewidth below 1 $MHz$.

\begin{figure*}[ptb]
	\centering\includegraphics[width=12cm]{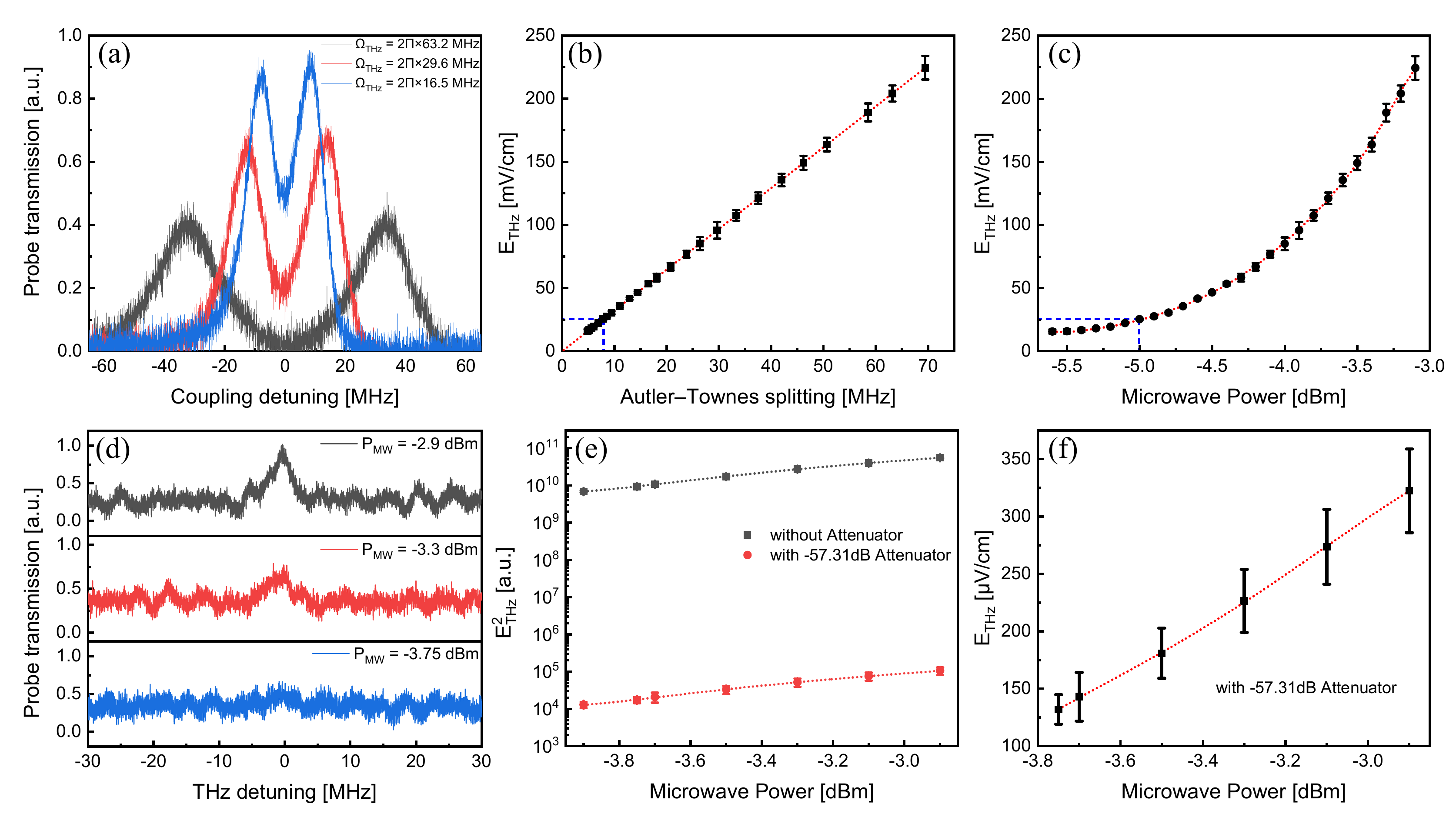}
	\caption{\label{fig2} Characterize the minimum detectable THz electric field.
		(a) The AT splitting of the $25S_{1/2}$ to $25P_{3/2}$ Rydberg transition that occurs for larger THz electric field with Rabi frequency $\Omega_{THz}$. By scanning coupling laser detuning, the peak splitting of probe transmission corresponds to $\Omega_{THz}$.
		(b) The linear dependence of THz electric field strength on the AT splitting. The blue dashed line marks where the AT splitting is equal to the EIT linewidth.
		(c) The THz wave electric field strength versus power of fundamental microwave.
		(d) The nonlinear region measurement of THz electric field with a minimum value $132\pm 13\mu V/cm$ (10 s detection). The probe transmission is observed by scanning the THz detuning.
		(e) The THz electric field in the nonlinear region is obtained precisely. The -57.31dB attenuator is used to reduce the THz eletric field from the accurately measured linear region.
		(f) The THz wave electric field strength in the nonlinear region versus power of fundamental microwave. The error bars in (b), (c), (e) and (f) indicate standard error from five measurements.}
\end{figure*}

\section{Calibrate the minimum detectable Terahertz electric field}
The transmission of the probe laser is measured when the coupling laser is scanned. With the THz field on resonance, the EIT peak splits due to the ATS as shown in Fig.~\ref{fig2}(a). The AT peak splitting is proportional to the Rabi frequency $\Omega _{{\rm{THz}}}$ of the THz electric field. With the calculated transition dipole moment $\mu _{THz}  =242 ea_0$, the electric field strength $E_{THz}  = \hbar \Omega _{THz} /\mu _{THz}$. This relation (linear region) is valid until the splitting is smaller than the linewidth of the EIT peak which is 7.69 $MHz$ in our experiment. To calibrate the minimum detectable THz field, the technique similar to the recently published paper \cite{ChenOpt2022} is applied.

In the linear region, the measurement of THz electric field is traceable and self-calibrated. The minimal detectable THz electric field in the linear region is $25.37\pm 0.77{\rm{mV}}/cm$. The electric field strength of THz wave at the atoms is firstly measured in the linear region shown in Fig.~\ref{fig2}(b). Then the relationship between THz electric field and fundamental microwave power in the linear region is calibrated shown in Fig.~\ref{fig2}(c). Our result shows that the output power of the THz source (32-times frequency multiplier chain) has a nonlinear relationship with the microwave power.

To generate the weak THz electric field, the -57.31 $dB$ attenuator is used. The attenuation value can be determined
directly from the measured two THz electric field strengths with and without this attenuator. After attenuation, the splitting is hard to be distinguished (nonlinear region). The THz detuning is swept to observe the peak of probe transmission (Fig.~\ref{fig2}(d)) whose value is positive correlation with THz field strength. From the precisely measured curve of THz electric field versus microwave power, the attenuated field strength can be determined accurately shown in Fig.~\ref{fig2}(e). In Fig.~\ref{fig2}(f), the variation of THz electric field with microwave power is measured in the nonlinear region. When the signal to noise ratios (SNR) equals 1, the minimum detectable THz electric field is $132\pm 13\mu V/cm$ with detection time duration 10 s).

\section{Rydberg-atom receiver for Terahertz communications}
To demonstrate the wireless THz communications using the atomic receiver, the protocol that imposes the modulated THz signal into transmission of probe beam when the the frequencies of laser beams are locked is performed \cite{MeyerAPL2018}. The digital signal of communications information is encoded in 8 states of modulation phase $\phi {}_{THz}$ generated by the digital synthesis signal generator (DDS) in Fig.~\ref{fig1}. For frequency modulation, this digital signal is directly sent to the microwave source. Meanwhile, an external microwave switch controlled by the digital signal is needed to implement the amplitude modulation. Through a 32-times frequency multiplier chain, the modulated THz wave is produced from the microwave. When the THz wave acts on the atoms, the transmission of probe beam is modulated via the EIT process. Thus an amplified balanced photodetector (PDB) is used to monitor the intensity of probe beam after the Cs vapor cell. The communications signal from PDB is analyzed and quadrature demodulated by a lock-in amplifier with reference signal from DDS. After that, the modulated probe transmission by the THz field is demodulated into an In-Phase voltage $V_I$ and a Quadrature-Phase voltage $V_Q$ with the phase $\phi _{de}  = \arctan (V_Q /V_I )$. The demodulated phase $\phi _{de}$ is corresponding to the modulation phase $\phi {}_{THz}$ of THz field.

\begin{figure*}[ptb]
	\centering\includegraphics[width=12cm]{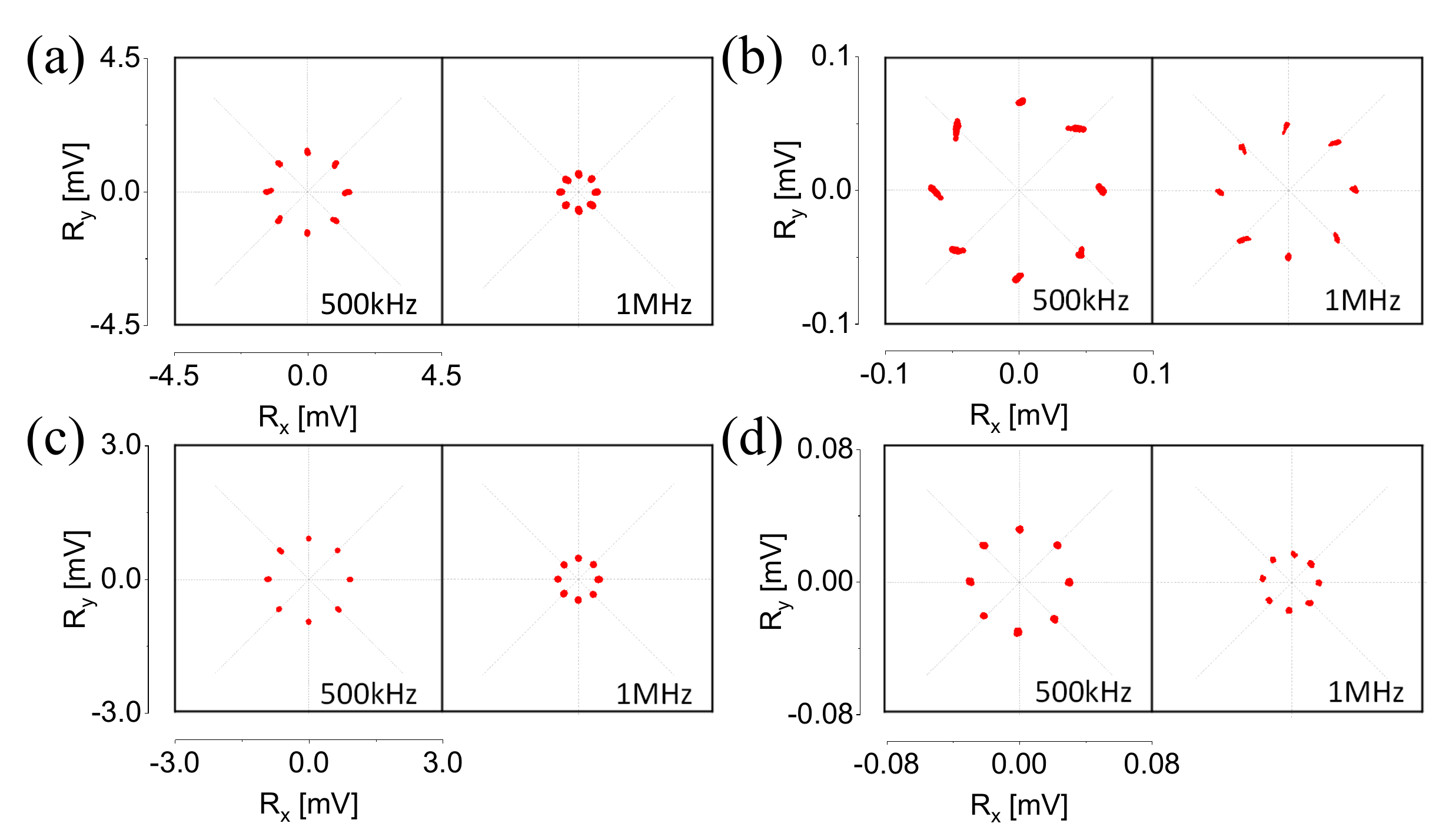}
	\caption{\label{fig3} The demodulated 8PSK diagrams of amplitude-modulated (AM) and frequency-modulated (FM) THz communications.
		To demonstrate the wireless THz digital communications using the atomic receiver, the protocols that impose the amplitude-modulated and frequency-modulated THz signals into transmission of probe beam are performed. The signals from PDB are analyzed and quadrature demodulated by a lock-in amplifier to obtain the 8PSK diagrams. The THz electric field is $E_{THz}=85.18mV/cm$ in the linear region and $E_{THz}=4.05mV/cm$ in the nonlinear region.
		(a) Amplitude modulation with THz electric field in the linear region. The amplitude of demodulation signal decreases when the modulation frequency becomes larger.
		(b) Amplitude modulation with THz electric field in the nonlinear region.
		(c) Frequency modulation with THz electric field in the linear region.
		(d) Frequency modulation with THz electric field in the nonlinear region.}
\end{figure*}

First, the amplitude modulation protocol is tried in the THz wireless communications demonstration. The THz wave is turned on and off by external microwave switch controlled by the phase modulated signal from DDS. In the linear region, the THz electric field is $E_{THz}=85.18mV/cm$ and the demodulated 8PSK constellation diagrams is shown in Fig.~\ref{fig3}(a). In the nonlinear region, the THz electric field is $E_{THz}=4.05mV/cm$ and the demodulated 8PSK constellation diagrams is shown in Fig.~\ref{fig3}(b). The 8 phase states are encoded from $\phi_{m}=0^{\circ}$ to $\phi_{m}=315^{\circ}$ with interval $45^{\circ}$. And the modulation frequency 500 $kHz$ and 1 $MHz$ is employed which is limited by the bandwidth of our PDB and lock-in amplifier. Both in the linear region and nonlinear region, the 8 phase states can be clearly discriminated. But the phase noise becomes lager in the nonlinear region since the THz field is weaker.

It's well known that frequency modulation can avoid the influence of amplitude interference in the propagation progress that can significantly improve the communications quality. Therefore, the frequency-modulated methods are also tried in our THz wireless communications scheme. In the frequency modulated communications, the fundamental microwave frequency hopping is controlled by the phase modulated signal from DDS. After 32-times frequency multiplication, detuning of THz wave hops from resonance to a large detuning 156.8 $MHz$. Relatively, transmission of probe beam varies from maximum peak to small value or even zero when the THz detuning jumps from resonance to 156.8 $MHz$. So the periodic frequency hopping can be converted into the intensity oscillation of probe beam passing through the Cs vapor cell. Then the collected signal of probe beam intensity by the PDB is sent to the lock-in amplifier. After quadrature demodulation, the 8PSK constellation diagrams illustrated in Fig.~\ref{fig3}(c)(d) can be acquired. From Fig.~\ref{fig3}, the atomic receiver has the same performance compared with amplitude modulation.

\section{Discussion}
The THz transmission distance is one of the biggest challenges for wireless communications as the serious propagation attenuation. For distance below 1 $km$, the free space path loss (FSPL) is dominated. The FSPL can be given in terms of the frequency as \cite{Pozar2011},
\begin{center}
\begin{equation}
	\label{FSPL}
	{\rm{FSPL}}=(\frac{4\pi df}{c})^2.
\end{equation}
\end{center}
In our system, the frequency $f=$ 338.7 $GHz$. Considering the distance $d=$ 1 $km$ and vacuum speed of light $c$, the FSPL is 143 $dB$. For distance large than 1 $km$, the atmospheric attenuation must be taken into account especially for THz wireless links with large carrier frequency. For example, the atmospheric attenuation is 5 $dB/km$ at 338.7 GHz \cite{NagatsumaNPon2016}. One way to extend the communications distance is to raise the power of THz source. However, this is very difficult for both photonic and electronic based THz source. Another way is to use the THz receiver with higher sensitivity, that's why we tried the Rydberg-atom receiver. In our THz wireless link, the output power of THz source corresponding to the minimum detectable THz electric field in the nonlinear region is -56 $dBm$. The full power of the commercial source we used can be 12 $dBm$, so there is link budget 68 $dB$ for propagation loss. The maximum transmission distance 3.5 $m$ can be achieved. Unlike the electronic receivers, the ultimate sensitivity of the atomic receiver is restricted by the standard quantum limit (SQL). The sensitivity for SQL is described as \cite{HaoquanJPB2015},
\begin{center}
\begin{equation}
	\label{sensitivitySQL}
	\frac{{E_{\min } }}{{\sqrt {Hz} }} = \frac{{2\pi \hbar }}{{\mu _{THz} \sqrt {NT_2 } }}.
\end{equation}
\end{center}

In our experiment, the atom number participating in the EIT progress is $N = 1 \times 10^5$. Considering dephasing time ${T_2 }=21 \mu s$, the SQL sensitivity is about 2.3 $nV{cm}^{ - 1} Hz^{ - 1/2}$. For room-temperature atoms, the effectiveness of EIT probing is reduced by 30-40 $dB$ \cite{MeyerPRA2021}. This means the total budget relative to the SQL for wireless transmission is 159-169 $dB$. Then corresponding distance is about 5 $km$. Recently, some excellent work has been carried out on embedding the Rydberg-atom sensor into antennas or waveguides \cite{SimonsIEEE2019,HollowayAPL2018}. If use a 50 $dBi$ transmitter antenna and embedding the Rydberg-atom sensor into a 50 $dBi$ receiver antenna, this distance will be over 18 $km$. In addition, the Rydberg transitions have large dipole moments in a wide frequency range from 100 $GHz$ to nearly 3 $THz$, so the atomic receiver can keep this performance in a very wide band \cite{WadeNP2017,MeyerJPB2020}. Unlike the bandwidth of electronic receiver which is limited by the antennas and waveguides, atomic receiver bandwidth depends on the Rydberg transitions and lasers.

For any type of communications receiver, the maximum channel capacity $C$ is important and described by the Shannon-Hartley Theorem\cite{MeyerAPL2018,CoxPRL2018},
\begin{center}
\begin{equation}
	\label{shannon}
	C=f_{m}log_{2}(1+SNR).
\end{equation}
\end{center}
Here the $f_{m}$ is the symbol frequency and the SNR is signal-to-noise ratio. The modulation frequency is limited by the time needed for the EIT/AT to develop \cite{MeyerAPL2018,CoxPRL2018,SimonsAPL2019}. Considering maximum 10 $MHz$ modulation frequency, the channel capacity is 10 $Mbits/s$ at the minimum detectable THz field.

Finally, the atomic receiver can be used in the THz wireless-to-optical link because it's able to directly convert the wireless signals into the optical ones \cite{DebAPL2018}. Usually the wireless-to-optical receiver systems rely on electronic devices. The signal of electromagnetic wave is received and mixed down to lower frequency. Then an electro-optic modulator can be used to convert this signal into optical domain \cite{SalaminNP2018,UmmethalaNP2019}. In the demonstration of our THz wireless communications system, the amplitude-modulated or frequency-modulated information is directly converted into laser signal at 852 nm by the Rydberg-atom receiver. Comparing with electronic ones, the wireless-to-optical atomic receiver is simpler and lower in cost without frequency down-mixing and extra modulator.

\section{Conclusion}
To summarize, the room temperature cesium Rydberg atoms are demonstrated to be sensitive THz electric field sensors. The minimal detectable THz electric fields $132\pm 13\mu V/cm$ (10 s detection) are calibrated. Furthermore, the phase-sensitive conversion of amplitude-modulated and frequency-modulated THz wave into optical signals is conducted to realize the communications with 8-state phase-shift-keying. The atomic receiver has potential to realize 18 $km$ terahertz wireless communications at 338.7 $GHz$. Furthermore, the atomic receiver can be used in the terahertz wireless-fiber link to directly convert the wireless signals into optical signals.

\begin{acknowledgments}
The work was supported by the Key-Area Research and Development Program of GuangDong Province (Grants No. 2019B030330001 and No. 2020B0301030008), the National Natural Science Foundation of China (Grants No. 61875060, and No. U20A2074), the Key Project of Science and Technology of Guangzhou (Grant No. 2019050001), the Natural Science Foundation of Guangdong Province (Grants No.2018A030313342), the Science and Technology Program of Guangzhou (Grant No. 202201010486), and the Guangdong Basic and Applied Basic Research Foundation (Grants No. 2022A1515012026).
\end{acknowledgments}


\end{document}